# Toward Automated Clinical Transcriptions

Mitchell A. Klusty, BS[1], W. Vaiden Logan, BS[1], Samuel E. Armstrong, MS[1], Aaron D. Mullen, BS[1], Caroline N. Leach, BS[1], Jeff Talbert, PhD[1], V. K. Cody Bumgardner, PhD[1]
[1]University of Kentucky, Lexington, KY, USA

**Abstract**

*Administrative documentation is a major driver of rising healthcare costs and is linked to adverse outcomes, including physician burnout and diminished quality of care. This paper introduces a secure system that applies recent advancements in speech-to-text transcription and speaker-labeling (diarization) to patient-provider conversations. This system is optimized to produce accurate transcriptions and highlight potential errors to promote rapid human verification, further reducing the necessary manual effort. Applied to over 40 hours of simulated conversations, this system offers a promising foundation for automating clinical transcriptions.*

**Introduction**

Accurate and timely documentation is essential in the healthcare sector, but manual transcription of patient-physician interactions is laborious, and errors are common. The extensive burden of documentation placed on clinicians takes away valuable time from patient care. Existing automated transcription systems offer a solution but can often struggle with the complexity of medical conversations, noisy environments, overlapping dialog, and differentiating between speakers.

This paper presents a system designed to significantly reduce the effort of manual transcription while ensuring the confidentiality and integrity of medical data. Integral to this system are two open-source tools, PyAnnote and OpenAI's Whisper. PyAnnote[1] enables speaker diarization, the segmentation of audio streams and identification of individual speakers. OpenAI's Whisper[2] generates accurate transcripts of audio streams, handling loud settings and overlapping speech. These models revolutionize the way transcriptions are performed. Our system provides a secure, efficient pipeline to merge them, creating one unified, speaker-labeled transcript of an audio or video file that is optimized for fast human verification. The user-driven automated analysis of files allows adjustments that create the most refined transcript for the various circumstances in each file, expediting any required manual cleanup.

Considerations of security are paramount when designing a system that processes sensitive medical information. Using general, commercially available models does not guarantee the security of such critical data. Our system implements robust security features including end-to-end encryption of data in motion, and encryption of data at rest. All analysis of data with Whisper and PyAnnote is performed on locally hosted, secured servers. This allows the system to be self-contained, avoiding any potential leaks of confidential data. HIPAA deployable instances of this system can be easily created on compliant servers to ensure necessary security protocols are followed.

The proposed system provides a scalable, automated and secure approach to generating transcriptions. It addresses operational and administrative challenges of clinical documentation by minimizing the normally cumbersome work of manually transcribing and embedding strong security measures at each component of the system. The result is faster documentation with fewer errors, as the bulk of the effort of creating an original, speaker-labeled transcript is already completed, so human effort can be put toward ensuring accuracy. This reduces the cognitive load on physicians and allows them to spend more time where it is most valuable, helping patients.

---

**Methods**
The developed system is composed of a simultaneous transcription and diarization pipeline, ending with a merging of the two, reconciling differences in timestamps to create a single speaker-labeled transcript. Figure 1 shows a high-level overview of the components of our system and how they work together. Details for each component will be discussed below.

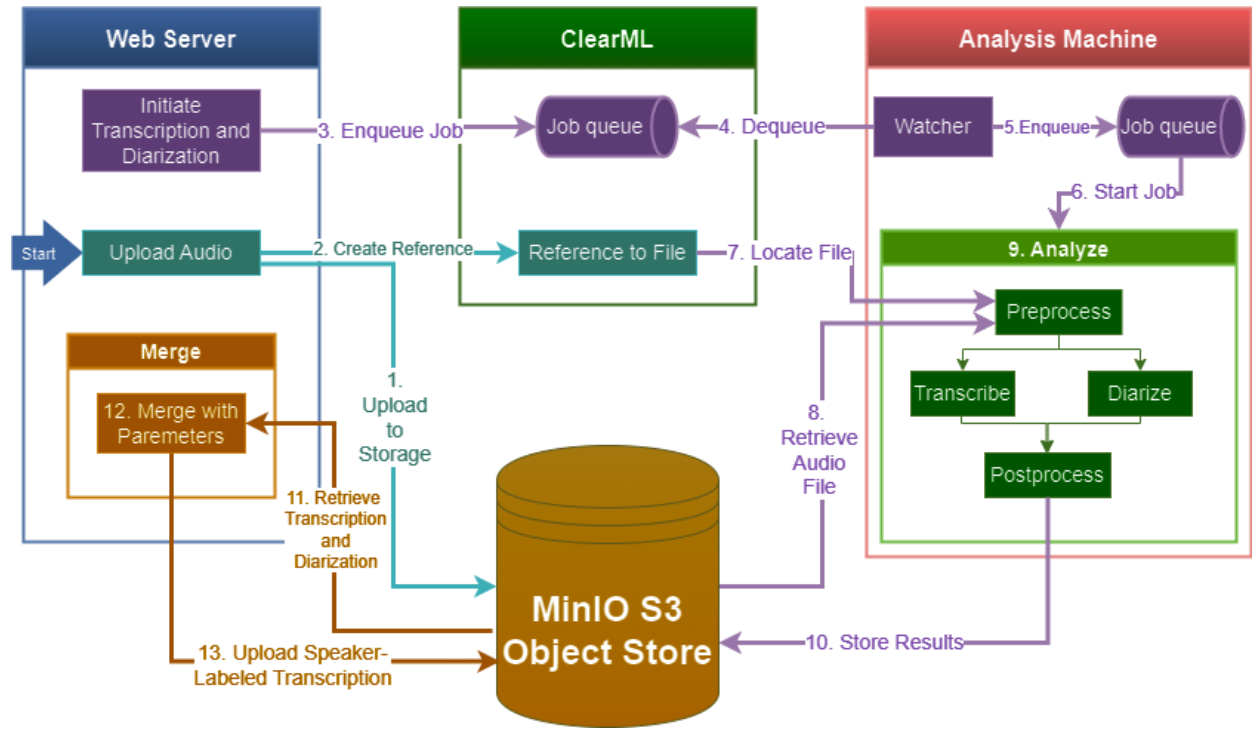

*Figure 1 Diagram of the full system showing how each individual component interacts*

1. **Audio Upload and Storage**
   The process begins with users uploading audio files though a web-based interface. These files are transferred to a locally hosted MinIO S3[3] server, which provides a secure and scalable storage solution. MinIO is an object storage system that is compatible with the Amazon S3 programming interface. S3 is typically seen in cloud environments, but by hosting the MinIO server locally, we maintain full control over how and where the data is stored. In medical contexts, assuring sensitive audio files remain within our secure system's infrastructure is crucial. Eliminating the need for external cloud providers also reduces the risk of data breaches on those services. A reference to this file, identifying its S3 address, is created within the ClearML system discussed in the next step that makes the file accessible to the analysis server (see Figure 1, steps 7 and 8), where transcription and diarization are performed in parallel. A HIPAA-compliant encryption and authentication policy is maintained to uphold the confidentiality and integrity of the data. Utilizing S3 for file storage allows for secure management of the files and guarantees they are directly accessible to the components that need them, and not stored where they are not needed.

2. **Job Scheduling and Management**
   Once the files are uploaded, the user can initiate transcription and diarization jobs through the web interface. Our system utilizes the ClearML[4] platform to manage the job queue and monitor ongoing jobs. ClearML serves as the link between the analysis server, the web interface, and the storage structure facilitating interaction with the processes and their results.

   When the user initiates the job through the web interface, it is queued on ClearML. A queue watcher program on the analysis server monitors ClearML, dequeuing jobs and adding them to the server's internal queue. This enables the system to support the execution of jobs in large batches, beginning analysis as soon as resources become available. ClearML provides the parameters of the job to the analysis server, including the location of the S3 address associated with the audio file stored in MinIO. The analysis server then retrieves the file and performs transcription and diarization. During runtime, logs from the process are output to ClearML, allowing the user to monitor the job's progression. Once it is completed, the finished transcription and diarization files are uploaded to a designated S3 location, and all files and artifacts are cleaned from the system.

This distribution of responsibilities ensures that each component—the web interface, MinIO object storage, and analysis server—remains scalable and independent. By decoupling these systems, each can be scaled independently based on demand. As necessary, additional web servers or resources can be added to allow more user requests. MinIO storage can be expanded to accommodate larger volumes of uploaded data, and additional computational power can be provided to the analysis server to increase speed of the analysis. Additionally, these components can be easily distributed to any server that can support them. This modularity allows for efficient allocation of available resources and allows the system to be adaptable for various workload sizes, performance requirements, and future improvements.

3. **Transcription and Diarization**

Transcription and diarization are performed simultaneously to reduce the total execution time. Transcription is performed using the powerful Whisper model. Whisper can identify words and phrases in the audio stream and report the timestamps at which they occurred, utilizing the surrounding context to increase accuracy. Whisper is also capable of identifying and grouping overlapping phrases said at the same time. The reported transcription is stored in a structured format tracking the overall conversation, each identified phrase, with timestamps for the start and end of the phrase and for each word in the phrase. It also tracks probabilities that each word was transcribed correctly.

The diarization is performed with the PyAnnote speaker-diarization-3.1[5] model. Through the web interface, the user can allow the model to identify the number of speakers in the file, or manually specify that number, preventing the model from making incorrect assumptions. PyAnnote supports overlap detection, permitting the system to identify when more than one speaker is talking simultaneously. The output of the diarization is a file that contains a set of time intervals and an ID representing the identified speaker at that time.

4. **Merging the Transcription and Diarization**

The main challenge emerges when combining the outputs of the two processes: the timestamps produced by the diarization do not match those produced by the transcription. To resolve this, our system treats the Whisper transcription as a source of record and attempts to label the speaker of each phrase. To accomplish this, the system matches the timestamps of each phrase with the probability of each speaker being the one who said it. This probability is calculated by finding the ratio of overlap between the transcription timestamps and the diarization timestamps for each speaker. For example, as seen in Figure 2, if 50% of the transcription time frame is covered by Speaker 1 and 80% is overlapped by Speaker2 in the diarization, the probabilities are calculated as:

$$Speaker\ 1\ probability = \frac{0.5}{0.5 + 0.8} \cong 0.385$$

$$Speaker\ 2\ probability = \frac{0.8}{0.5 + 0.8} \cong 0.615$$

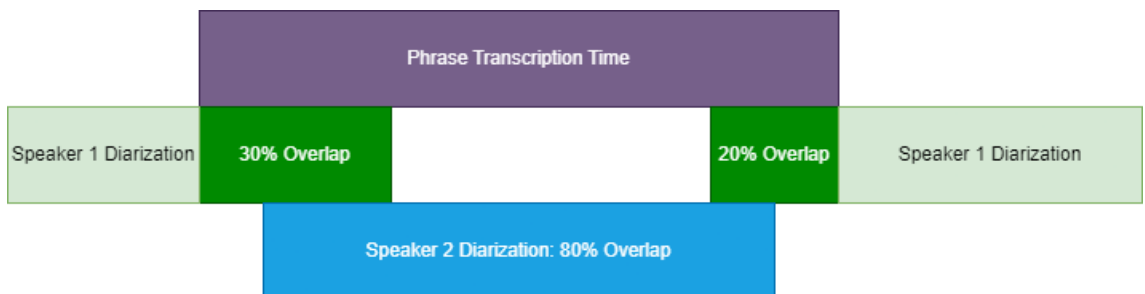

*Figure 2 Example showing calculation of transcription-diarization overlap to predict the speaker*

This approach attempts to mesh the often-disparate timestamps of the translation and diarization. It does not, however, work consistently well on all audio files. We found that files with lower audio quality and more overlap in dialog performed worse with this method, so an additional metric was developed. Roles/names can be assigned dynamically for each speaker in the file, and a large language mode (LLM) will comb through the transcription to determine, with contextual awareness of the dialog, which of the speakers is most likely to have spoken that line. This guess is then weighted with a user-defined value to determine how strong of an effect it will have on the probabilities. In the above example, if the LLM responded that, contextually, Speaker 2 was more likely the speaker, and the weight for its response was .25 (25%), the probabilities would be recalculated as:

$$Speaker\ 1\ probability = 0.75 * \left(\frac{0.5}{0.5 + 0.8}\right) + 0.25(0) \cong 0.288$$

$$Speaker\ 2\ probability = 0.75 \left(\frac{0.8}{0.5 + 0.8}\right) + 0.25(1) \cong 0.712$$

This augmentation works particularly well for audio files where the transcription is accurate and the diarization model struggles to properly label the speakers. In cases like this, we found that a higher weight on the LLM's decisions provided more accurate diarization. It works less well when the diarization from PyAnnote is more successful because it adds a needless complication to the probabilities, attempting to address issues that don't exist. It also performs worse when the transcription performs worse. If the transcription hallucinated lines not in the text or did not properly add lines, this interrupts the flow of the dialog and confuses the LLM as to which speaker contextually said each properly retrieved phrase. In situations like these, a weaker weighting should be placed on the LLM to minimize mislabeling.

Because the merging step can be different between every file, the user can easily adjust the labels for each speaker and the weight applied to the LLM's decisions and quickly rerun the process. The result is a speaker-labeled transcription with high enough accuracy to ensure any necessary human revisions are simple and easy.

5. **Scoring the System**

To measure the performance of the system overall, we used a dataset[6] containing simulated patient-physician medical interviews. This dataset was chosen because it mimics real-world scenarios with distinct speakers and varying speech patterns, allowing for strong evaluation of the transcription and speaker diarization. This allowed us to validate the system in a controlled and simulated yet realistic setting.

229 files were transcribed and diarized, then, the outputs were merged as a batch. This amounts to nearly 44 hours of audio and more than 300,000 words transcribed. The dataset was divided into 5 domains: Gastrointestinal, Musculoskeletal, Cardiac, Dermatological, and Respiratory.

The labels of *Doctor* and *Patient* and a weight of 0.45 were placed on the LLM's decisions. In practice, the weight on the LLM will generally be in the range 0.0 when the diarization is ideal to 0.6 when the diarization performed very poorly. This range allows the LLM to make up for the faults of the diarization model without ignoring its findings. A small sample was taken of the processed files, and 0.45 was the weight that performed the best on that sample. As discussed above, generally, the LLM weight should be adjusted per file to make the merge calculations specific to that file. The weight of 0.45 was standardized to all files to score them consistently and demonstrate the need for individual treatment of each file. We then compared the transcript in the dataset with our generated transcription. Both transcriptions were normalized, converting some of the common words found in the transcripts that were both synonyms and homophones but were spelled differently between the transcripts (i.e.: "ok" and "okay", "uhm" and "um", "titres" and "titers"). Words were also adjusted to be case insensitive and remove punctuation. We then calculated the number of words that were transcribed correctly, the number in the original script that were not found by our system, the number that were hallucinated by our system (generated but did not exist in the original transcript), and the number of words that replaced a word in the original transcript. These numbers have been analyzed in the results below.

We scored the Whisper transcription and our LLM-augmented PyAnnote speaker diarization separately as, though the speaker labeling is somewhat dependent on the accuracy of the transcriptions, they are separate metrics. To standardize the scoring across all files, the transcripts provided in the dataset (which we refer to as original text or original transcription) are treated as a gold-standard, accurate transcription of the audio. This is despite known errors in those original texts, such as misspellings, and simple incorrect transcriptions. To measure the accuracy of the transcriptions, we use the distribution of Word Error Rates (WER). This is a measurement of accuracy of the transcriptions. It is calculated by adding the number of missed words, hallucinated words, and replaced words and dividing this by the total number of words in the original text. Therefore, a lower WER value indicates a more accurate transcription. We then calculated the median WER for each of the 5 domains in the dataset to find possible trends in the capabilities of Whisper across various medical topics. We also calculated the median percentages of each of the measured categories: Correct (properly transcribed), Missed (not transcribed), Hallucinated (transcribed but were not in the audio), and Replaced (words in the audio that were improperly transcribed). To measure the speaker labeling, we calculated the number of Correct and Replaced words in the generated transcript that were correctly labeled and generated a score by dividing this number by the total number of words in the original text. Again, we found the distribution of these scores and the average percentages of mislabeled words for each domain.

**Results**

Figure 3 shows the distribution of WER for each of the audio files. The overall median WER was 0.145, and Figure 3 shows that most transcriptions achieved a WER under 0.2. This indicates good performance, as generally less than 20% of words caused errors in the generated transcriptions.

Figure 4 shows the distribution of percentages of mislabeled words in each file. The overall median is 0.233, but the distribution has a wide range, with some texts containing >80% mislabeled speakers. This highlights the importance of adjusting the merge step for individual files to achieve the best speaker-labeled transcripts. We found the Whisper transcriptions would occasionally add or remove words from the audio, and, in rare instances, would completely ignore several sentences, interrupting the transcription. These errors would bleed into the speaker labeling, as that process depends on the accuracy of the transcription. These errors are more pronounced with a higher weight on the LLM's decisions. Breaks in the dialog interrupt the flow of the LLM's understanding of the recording, making accurate labeling harder. In practice, a file that is found to have a high speaker mislabeling rate should be rerun with adjusted settings, and, ideally, a human-corrected transcript to find better results.

Figure 5 displays the average makeup of both the original text and generated transcription. It shows the percentages of the words from the original text that were correctly transcribed, not transcribed, or transcribed differently when compared to the generated transcription. It also shows the percentages of words in the generated transcription that matched the original text, the percentage that were added that did not exist in the original text, and the percentage that were transcribed differently than the original.

Figure 6 shows for each medical domain the percentage of words from the original text that were untranscribed or replaced, as well as the percentage of words from the generated text that were hallucinated or mislabeled. Most of these values are roughly consistent between domains, except for the percentages of speaker-mislabeled text, but we could not find any correlation between the domain and the

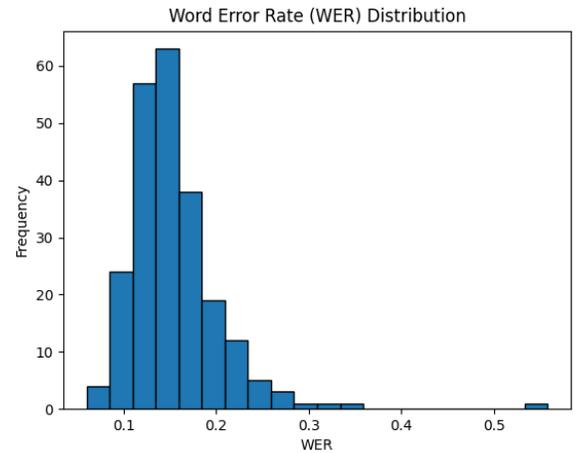

*Figure 3 A graph showing the distributions of Word Error Rates*

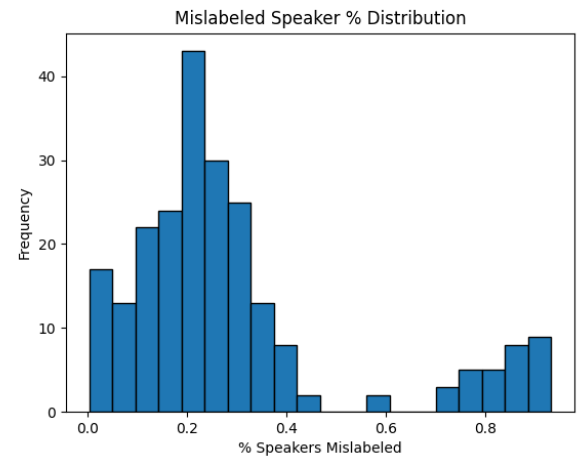

*Figure 4 A graph showing the distributions of mislabeled speakers*

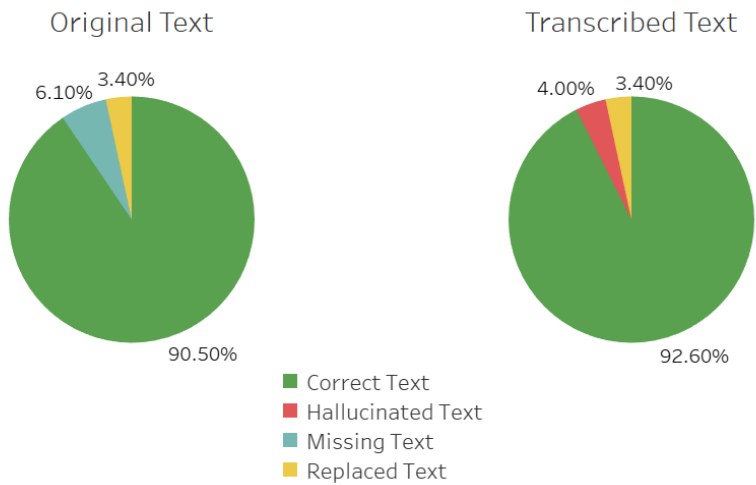

*Figure 5* Pie charts detailing the percentages of words in the original text and transcribed text

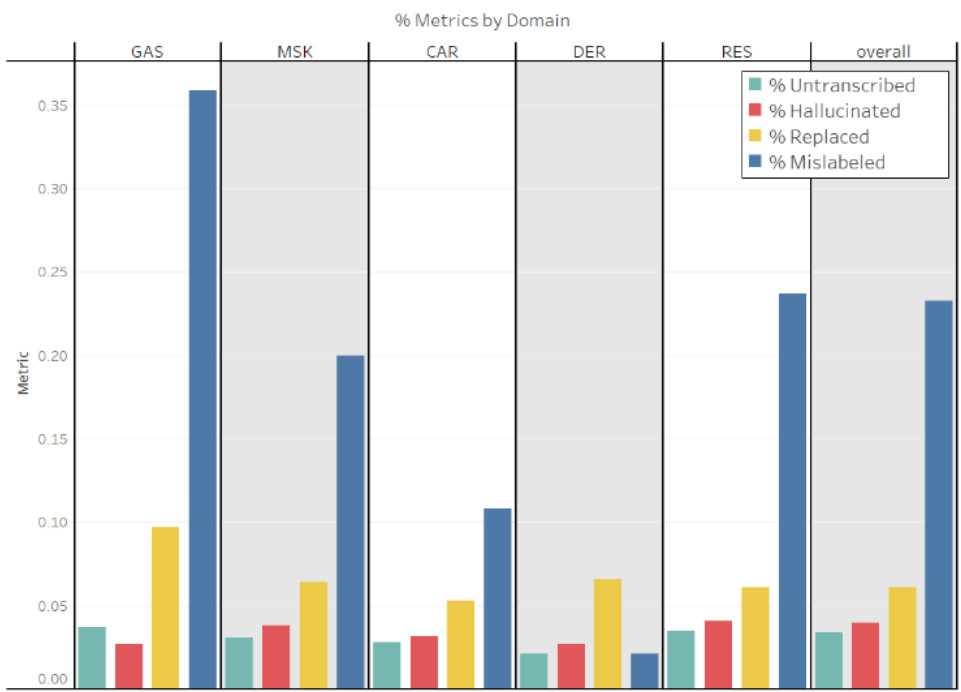

*Figure 6* Bar graph showing the breakdown of incorrect words in the transcription and the percentage of words with the speaker mislabeled, separated by domain

**Discussion**

The developed system creates accurate transcriptions on the first pass. Leveraging the advanced Whisper transcription and PyAnnote speaker-diarization models, it generates high-quality, speaker-labeled transcripts that correctly capture the majority of dialog and speaker distinctions. While there can be errors in the output, the system is designed to facilitate easy human revisions to finalize the definitive transcription. Our system performs most of the effort, creating transcripts which significantly reduce the required human processing time associated with manual transcription. The system outputs a transcript grouped by phrase with associated confidence scores reflecting the models' accuracy both in transcription and speaker-labeling. This makes for efficient identification of potentially incorrect elements and minimizes the effort to correct errors.

This system was designed with security in mind. Industry standard end-to-end encryption is maintained for all data in transmission, and MinIO and other S3 alternatives support at-rest encryption of stored files. The division of responsibility across components of the system allows for each of them to be deployed independently with the necessary resources and security protocols, isolating any compromise or system failure.

**Conclusion**

This paper introduced a system which integrates cutting-edge AI models to create speaker-labeled transcriptions of audio/video files, providing a significant step forward in automating clinical documentation. Combining Whisper for transcription and PyAnnote for speaker diarization, the system achieved high-quality, speaker-labeled transcriptions of a dataset with over 40 hours of simulated patient-physician conversations.

There are limitations, however, in that accurate speaker-labeling is dependent on accurate transcriptions, making the system not fully autonomous. To guarantee the highest fidelity of the transcription, human verification is essential. Our system addresses this by supporting LLM-augmented user adjustments to ensure the highest accuracy of a first pass output and flags any likely errors to expedite revisions and make a fully accurate transcript. Despite these limitations, our system provides a strong foundation for future improvements in medical transcription automation.

The ability to store sensitive data securely with MinIO, combined with ClearML's job management services, makes the system extremely modular. This modularity creates a secure, scalable, flexible structure that enables quick deployment.

As administrative overhead in healthcare continues to grow, this could reduce time spent on clinical documentation by creating reliable, human-verified transcripts in significantly shorter times. Future work will focus on further improvements to speaker identification and automation while maintaining accuracy standards and securely connecting recording devices to this system to allow for immediate documentation.